\documentstyle[12pt,titlepage]{article}
\newcommand {\ks} {km~s$^{-1}$}
\newcommand {\h} {$\ h^{-1} \, Mpc \;$}
\def\ref{\par\noindent\hangindent=1cm}
\begin{document}
\title{}
\author{}
\date{}
\vspace{4cm}
\Huge
\begin{center}
{\bf Optical Radii \\
of Galaxy Clusters }
\end{center}
\vspace{2.5cm}
\normalsize
\begin{center}
{\bf M. Girardi$^{1,2}$, A. Biviano$^{3}$,\\
G. Giuricin$^{1,2}$, F. Mardirossian$^{1,2}$, and M. Mezzetti$^{1,2}$}
\end{center}
\vspace{0.5cm}
(1) SISSA, via Beirut 4, 34013 - Trieste, Italy \\
(2) Dipartimento di Astronomia,
Universit\`{a} degli Studi di Trieste, Trieste, Italy \\
(3) Institut d'Astrophysique de Paris, 98bis bd Arago, 75014 Paris, France \\

$\ $

$\ $

\noindent
E-mail:

\noindent
girardi@tsmi19.sissa.it;\\
biviano@iap.fr;\\
giuricin@tsmi19.sissa.it;\\
mardirossian@tsmi19.sissa.it;\\
mezzetti@tsmi19.sissa.it

\pagebreak
\renewcommand{\thesection}{\arabic{section}}
\renewcommand{\thesubsection}{\thesection.\arabic{subsection}}
\section*{Abstract}
We analyze the density profiles and virial radii for a sample of
90 nearby clusters,
using galaxies with available redshifts and positions. Each cluster has at
least 20 redshifts measured within an Abell radius, and all the results come
from galaxy sets of at least 20 members.

Most of the density profiles of our clusters are well fitted by
hydrostatic-isothermal-like profiles. The slopes we find for many cluster
density profiles are consistent with the hypothesis that the galaxies are in
equilibrium with the binding cluster potential.

The virial radii correlate with the core radii at a
very high significance level. The
observed relationship between the two size estimates is in agreement with the
theoretical one computed by using the median values of the density profile
parameters fitted on our clusters.

After correcting for incompleteness in our cluster sample, we provide the
universal distributions functions of core and virial radii (obtained within
half an Abell radius).

\vspace{1cm}

{\em Subject headings:} galaxies: clusters of -- galaxies: clustering
\pagebreak
\section{Introduction}
The distribution functions of observational cluster quantities, such as radii,
velocity dispersions, masses, luminosities and X-ray temperatures, can provide
strong constraints both on cosmological scenarios and on the
internal dynamics of
these systems. Theoretical as well as observational estimates of these
distribution functions are presently being debated.

In previous papers (Girardi et al. 1993, hereafter GBGMM; Biviano et al. 1993),
we have addressed the distributions
of cluster velocity dispersions and virial masses.
Here, we deal with the spatial
distribution of member galaxies in clusters. Existing relations between the
linear sizes and other physical quantities of bound systems of galaxies have
been investigated in the literature, both observationally and
theoretically (see, e.g., Oemler 1974: Bahcall 1975; Peebles 1976; Dressler
1978; Sarazin 1980; Giuricin, Mardirossian, \& Mezzetti 1984; West, Dekel, \&
Oemler 1987; West, Oemler, \& Dekel 1989; Peebles, Daly, \& Juskiewicz 1989;
Lilje, \& Lahav 1991; Cavaliere, Colafrancesco, \& Scaramella 1991; Edge, \&
Stewart 1991; Inagaki, Itoh, \& Saslaw 1992; Kashlinsky 1992; and references
therein).

Several size estimates are present in the literature: the core radius,
the effective radius, the virial radius, the harmonic radius, the gravitational
radius, the mean radius, etc. (see, e.g., Bahcall 1977, and references
therein). In the present paper we deal with virial and core radii.

Relaxation processes naturally lead to the formation of physical cores in
clusters of galaxies  (see, e.g., King 1966; Lynden-Bell 1967; Peebles 1970;
Sarazin 1986). Generally, in order to fit the galaxy distribution of clusters,
the following surface density profile has been assumed
(see, e.g., Bahcall 1977, Sarazin 1980, and references therein):
\begin{equation}
\sigma(r)=\frac{\sigma_0}{[1+(r/R_{c})^{2}]^{\alpha}}
\end{equation}
where $\sigma_{0}$ is the $central$ $projected$ $galaxy$ $density$ and $R_{c}$
is the $core$ $radius$.
This model corresponds for $r >> R_{c}$ to a $volume-density$
$\rho(r) \propto r^{-2\alpha-1}$.
Many authors (see, e.g., Bahcall 1975, Dressler 1978,
Sarazin 1986) have used the King (1962) model to fit the cluster profiles;
this corresponds to the
model  of Eq.(1) with $\alpha = 1$. For instance, Bahcall (1975)
found an average value of $King-R_{c} \simeq 0.12 \pm 0.02$\h
(the Hubble constant is $H_{0} = h \, 100 \, km \, s^{-1} \, Mpc^{-1}$)
on a sample of
15 rich regular clusters; on the contrary, Dressler (1978) found an average
value of $King-R_{c} \simeq 0.24 \pm 0.06$\h on another sample of 12 clusters.
{}From the large compilation by Sarazin (1986), it may be noted that very
different values of King $R_{c}$ for the same cluster have been obtained by
different authors.

Core-radii have also been derived for the X-ray surface brightness
distributions, by fitting these profiles to hydrostatic-isothermal models
(see, e.g., Abramopoulous \& Ku 1983; Jones \& Forman 1984;
Lubin \& Bahcall 1993; Bahcall \& Lubin 1993;
and references therein). An average value of $\sim
0.05$--$0.15$\h is suggested for X-ray core radii, although both smaller
(Durret et al. 1993) and larger (Jones \& Forman 1992) values have been found.

Attempts to evaluate the core-radius of the cluster dark matter component have
been made (see, e.g., Eyles et al. 1991; Gerbal et al. 1992;
Kneib et al. 1993; Miralda-Escud\'e 1992; Soucail 1992). The results are still
preliminary.

In this paper we use the hydrostatic-isothermal models in order to fit the
cluster density profiles. In particular, this choice is corroborated by the
recent results of Bahcall \& Lubin (1993; see also
Gerbal, Durret, Lachi\`eze-Rey 1994), who claim that the so-called
``$\beta-$discrepancy''
can be solved by fitting the galaxy  density profiles with
models having $\alpha = 0.7 \pm 0.1$ in Eq.(1), i.e. $\rho(r) \propto
r^{-2.4\pm0.2}$ at large values of $r$,  instead of using the classical {\em
King} models. We remind that the $\beta-$discrepancy is the difference
between the observed values of $\beta_{spec}$,
determined from cluster X-ray temperatures and velocity dispersions,
and $\beta_{fit}$, determined from the gas and the galaxy density profiles
in galaxy clusters (see, e.g., Sarazin 1986).

The {\em virial radius} has been extensively used in the literature to evaluate
the sizes and masses (via the Virial Theorem)  of galaxy systems (see,
e.g., Limber \& Mathews 1960; Smith 1980; Fernley \&
Bhavsar 1984; West, Oemler, \& Dekel 1989; Perea, del Olmo, \& Moles 1990;
Pisani et al. 1992; Biviano et al. 1993). In the present paper we consider the
2D-projected virial radius (i.e. not de-projected by the classical value of
$\pi/2$),
\begin{equation}
R_{v}=\frac{N^2}{{\sum^{N}_{1}}_{i\neq j} R_{ij}^{-1}},
\end{equation}
where $N$ is the number of cluster members and $R_{ij}$ is the projected
distance between the $i$ and the $j$ member.

In this framework, we deemed it interesting to obtain estimates of cluster
sizes using a large data-sample of 90 clusters, and we examine the
relationships between these sizes and other cluster properties. In \S~2 we
describe our data-sample; in \S~3 we consider the method of analysis; in \S~4
we provide the results and the relevant discussion; and in \S~5 we list our
conclusions.

\section{The Data-Sample}
We based our analysis on a sample of 90 nearby ($z < 0.15$) clusters, each
one with at least 20 measured redshifts of cluster members within 1.5\h.
In this compilation we included some poor clusters (or rich groups)
in order to enlarge our range of cluster-richness.

The problem of foreground and background contamination may be severe when
redshift information is not available for all the galaxies considered. In fact,
different corrections have been adopted in the literature, leading to very
different estimates for the sizes of the same clusters (see, e.g., Bahcall
1975; Dressler 1978; Fitchett \& Merritt 1988; West, Dekel, \& Oemler 1987).

On the contrary, using observed redshifts and positions for all our cluster
galaxies, we have been able to assign the cluster membership. Therefore, our
sample should be considered quite free from contamination. The
procedure adopted to reject non-cluster members is similar to that described in
GBGMM. However, in this paper we have decided to adopt a
somewhat stricter selection criterion, which is described below.

In order to reject interlopers in the clusters, we make use of the distance of
a galaxy from the cluster center in the coordinates and in
the velocity space. The
center is defined as a biweight mean of the velocities, right ascensions and
declinations (see, e.g., Beers, Flynn, \& Gebhardt 1990, GBGMM).

A first selection is performed to separate obvious groups which are projected
in the same cluster field. First, we identify all groups separated by a
weighted gap larger than 4 in the velocity space.
A weighted gap in the space of the ordered velocities is defined as
the difference between two contiguous velocities, weighted according to
their rank in the ordered distributions: the closer to the center
of the distribution is the gap, the higher is its weight
(for further details see, e.g., Beers et al. 1990).
We then select as our cluster
the larger of these groups. Of the remaining galaxies, we subsequently reject
all of them lying outside a circle of 3\h from the cluster center.
This rejection is made in an iterative way, recomputing the center after each
iteration and thus redefining the 3\h circle and the rejected galaxies, until
convergence is reached.

This defines the main cluster body. Further selection requires the exclusion of
galaxies more distant than 4000\ks from the cluster mean velocity (as
in Beers et al. 1991, GBGMM),
and further rejection of galaxies separated from the
main cluster body by a weighted gap larger than 4 in the velocity space. The
greatest difference with respect to GBGMM is in the last selection in
the coordinate-space: we reject all galaxies more distant than 1.5\h from the
cluster center, once again in an iterative way. Since the size estimators are
not as robust as the estimators of velocity dispersions used in GBGMM,
we believe that restricting the analysis to galaxies within 1.5\h is
a very simple but efficient method to reduce the problem of interlopers,
since the higher the aperture the larger
the ratio between background/foreground galaxies and cluster members. Thus
it is more conservative to consider the inner cluster regions.

Problems of magnitude incompleteness may cause errors in the size estimates
(e.g., Dressler 1978). In particular, if a deeper sampling of galaxies (in
terms of magnitudes) has been made by observers in the central regions, the
shape of the density profile will look steeper than it is in reality. We have
therefore analysed all our clusters with available galaxy magnitudes, by
looking at the galaxy magnitudes as a function of clustercentric distance. We
have noted that some clusters are sampled more deeply
close to the center than they are outside.
We have rejected the fainter galaxies of these  clusters so as to
erase any trend of the limiting magnitude with clustercentric distance.

Yet another sort of bias may be present, if some regions in the cluster are not
covered observationally (generally, in the outer parts). As for the previous
case, the shape of the density profile will look artificially steeper than it
is in reality. There is no easy way to eliminate this sort of bias {\em a
priori,} but a signature of it may be the steepening of the density profile
with increasing apertures. As a consequence, we look with more confidence on
those results obtained from the central regions of our clusters, where problems
of incompleteness should not be so strong (see \S 3).

Yet another problem is the presence of substructures which may bias the
determination of cluster sizes, and in particular of the core radii (see, e.g.,
Sarazin 1980). When the cluster has several density clumps, one density profile
might not adequately fit the galaxy distribution; and large, spurious values of
the core radius are expected when a single profile is forced to fit all the
data. In order to investigate the possible presence of subclustering in our
sample, we performed the test of Dressler \& Schectman (1988a). This test gives
the probability,
$1-P_{DS}$, that some subclustering exists in the cluster (1000
simulations were run, according to the prescriptions of the above-mentioned
Authors). Actually this method (like others available in the literature) is
not efficient when the number of galaxies considered is low ($N < 40$). As a
consequence, we considered the subsample of our clusters, with $N \geq 40$
members and no evidence of subclustering, in order to check our results. As
detailed in \S 4, our results do not seem to be biased by subclustering.

The sample of the 90 clusters considered is listed in Table$\ $1, together with
the number $N$ of galaxy members inside 1.5\h from the cluster center, the
Abell counts $C$, the richness $R$, the Bautz \& Morgan and the Rood \& Sastry
classes (BM type and RS type), the probability
$P_{DS}$ that observed structures in the cluster are chance fluctuations
(i.e. low values of $P_{DS}$ indicate a high probability of subclustering),
and the relevant reference sources.

\section{The Method of Analysis}
In order to obtain the values of $R_{c}$, we fitted the observed galaxy
distributions to hydrostatic-isothermal models on the hypothesis of spherical
symmetry. Although clusters are not spherical symmetric (their flattening
may indeed be quite large, see, e.g., Rhee, van Haarlem, \& Katgert 1992),
we prefer not to include ellipticity as a free parameter in our
fits, because it would be poorly constrained in many clusters of our
data-set, where we do not have enough data, and because inhomogeneous
sampling during observing may result in artificial shapes for some clusters.
We nonetheless performed Montecarlo realizations of distributions
of 25, 50, 100 and 200 points extracted from King density profiles
with varying $R_{c}$ and ellipticity, and we analyzed how the derived values
of $R_{c}$ depend on the assumption of spherical symmetry. Not surprisingly,
these values are intermediate between the values of the major and minor
axis core-radii, and consistent with both within the formal errors.
Therefore, we think that the assumption of spherical
symmetry is a good working hypothesis.

The fitting was performed using the Maximum Likelihood technique which
takes directly into account the positions of individual galaxies, and does not
require any binning of the data (see, e.g., Sarazin 1980). This method is very
efficient, since it does not introduce any artificial scales, and it allows the
simultaneous fitting of $R_{c}$ and $\alpha$. On the basis of the discussion of
the previous section about interlopers, we assumed that the background galaxy
density was negligible, and we set it at zero.

We allowed  $R_{c}$ and $\alpha$ to vary from $0$ to $1$ and from $0.5$ to
$1.5$, respectively; we rejected values of $R_{c}$ larger than 1\h and
values of $\alpha$ outside the above mentioned interval. Moreover,  we did not
consider size-estimates for clusters whose fitted profiles are rejected by a
Kolmogorov-Smirnov test (hereafter KS test) at 90\% of probability. However,
fitting both $R_{c}$ and $\alpha$ at the same time is not an easy task,
especially when the data-set is poor. In a second stage of the investigation
(see \S 4), we therefore decided to fix the value of $\alpha$ at the median
value obtained in the two-parameter-fitting procedure. By reducing the free
parameters to one, $R_{c}$, we were able to better constrain the
size-estimates.

The errors on our $R_{c}$ were estimated by following the prescriptions of
Avni (1976), while the errors on our $R_{v}$ were obtained via the
Jacknife technique (see, e.g., Efron 1979, Geller 1984).

\section{Core and Virial Radii}
We obtained both the virial radii and the best fits of cluster profiles with
three increasing apertures, i.e. 0.5, 0.75, and 1\h.

In Table$\ $2 we give the virial radii for our sample of clusters; in col.(1)
we list the cluster name; in cols. (2), (3), and (4) the values of $R_{v,0.5}$,
$R_{v,0.75}$, $R_{v,1}$ and their errors, computed within the increasing
apertures 0.5, 0.75, 1\h for each cluster (if at least 20 members are
contained within the respective aperture).

In Table$\ $3 we give the median values of the core radii $R_{c,0.5}$,
$R_{c,0.75}$, $R_{c,1}$, and of the $\alpha$'s $\alpha_{0.5}$, $\alpha_{0.75}$
$\alpha_{1}$ for our clusters, computed within the apertures of 0.5, 0.75, and
1\h, respectively. The numbers of cluster fittings, the core radii, and the
values of $\alpha$ are listed in cols. (1), (2), and (3). The fittings were
performed only if at least 20 members are contained within the respective
aperture. Only the $good$ fits were considered (see \S 3).

The fitted profiles seem to be steeper when apertures larger than 0.50 $\
h^{-1} \, Mpc$ are considered. This trend may be explained either by sampling
incompleteness, or by a true difference between  the profiles of the inner and
the outer cluster regions (see, e.g. Bahcall \& Lubin 1993). When we restrict
our analysis to the inner regions, we note that the distributions of
$\alpha_{0.5}$ and $\alpha_{0.75}$ do not differ significantly, according to
KS- and Sign-test; the same holds for $R_{c,0.5}$ and of $R_{c,0.75}$. The
profiles seem to be well-defined within 0.75\h; therefore, we restricted our
analysis to this inner region.

Fixing the $\alpha$ parameter in the fitting procedure makes it
possible to reduce the
formal errors on the derived values of the core radii. Our median value of
$\alpha_{0.5}$ is $0.8^{+0.3}_{-0.1}$. It does not differ, within the errors,
from $0.7\pm0.1$, which is the $\alpha$-value suggested by Bahcall \& Lubin
(1993) (see also Bahcall 1977) to resolve the $\beta-$discrepancy for
clusters of galaxies.  So, we chose the value of $0.8$ in our
$\alpha$-fixed models.

In Table 4 we list the core radii obtained with the fixed $\alpha=0.8$
profiles; in col.(1) we list the cluster name; in col. (2) we list the
values of $R_{c,0.5,\alpha=0.8}$  and their errors,
computed within the apertures of 0.5\h, and in col. (3)
the corresponding confidence levels (in percent) $P_{0.5,\alpha=0.8}$
of the profile fittings (obtained by using the KS-test);
in col. (4) we list the
values of $R_{c,0.75,\alpha=0.8}$  and their errors,
computed within the apertures of 0.75\h, and in col. (3)
the corresponding confidence levels (in percent) $P_{0.75,\alpha=0.8}$
of the profile fittings (obtained by using the KS-test).

As usual, the  fits were performed only on samples with
at least 20 galaxies contained within the aperture considered.

The median values and the distributions of $R_{c,0.5,\alpha=0.8}$ and
$R_{c,0.75,\alpha=0.8}$ do not differ, according to the
KS- and Sign-test. This suggests that our estimates of the core-radii
are quite stable within half an Abell radius.

In order to detect a possible effect of subclustering on the radii
estimates,  we considered the subsample of our clusters with $N \geq 40$
members (for the reasons outlined in \S 2). On this sample we compared
the distributions of $R_c$ (and $R_v$) for clusters with and without
evidence of subclustering: there is no difference,  according to the KS-test.
We caution that this result cannot be considered neither
general, since our cluster sample is not
volume-complete, nor conclusive, since
different tests for subclustering may yield different results (see, e.g.
West \& Bothun 1990).
In fact, a detailed analysis of 16 well-sampled cluster (Escalera et al.
1994) has shown that the estimates of cluster
virial radii do decrease when galaxies belonging to
subclusters are identified and removed from the sample. However,
the changes are not statistically significant in most cases, being
important only when the cluster has a {\em bimodal} configuration,
and this happens for only 4 out of the 16 clusters considered,
In most cases the masses of the substructure are $\sim 10$ \% of their parent
cluster masses, so it is not surprising that subclustering does not
have a very large influence on the size determination.

The analysis of the errors of $R_{v}$ and  $R_{c}$ suggests that the dispersion
of the values of $R_{v}$ and  $R_{c}$ is partially intrinsic, and not mainly
induced by the errors involved.

\subsection{Core Radii vs. Virial Radii, and Other Relations}
Both the core and the virial radius describe the galaxy distribution inside the
cluster, so the existence of a relation between these two quantities is quite
natural. This functional dependence can be easily derived when one knows the
density profile.

On the hypothesis that the models we use fit cluster density distributions, we
obtain the following relation
\begin{equation} R_v = R_c\cdot F(A/R_c),
\end{equation}
where $A$ is the aperture considered. See the Appendix for  details.

In our data-sample the values obtained for the core and virial radii are very
well correlated, at $> 99.99$~\% significance level, for both the aperture of
0.5 and 0.75\h. Fig.~1 shows $\log R_{c,0.5,\alpha=0.8}$ vs. $\log
R_{v,0.5}$ (panel a), and $\log R_{c,0.75,\alpha=0.8}$ vs. $\log R_{v,0.75}$
(panel b). This correlation is not induced by obvious observational biases
(i.e., the radii do not correlate with the cluster distance, or with the number
of cluster members we use, etc.). These relations
remain practically unchanged if
clusters with significant subclustering are removed.  The theoretical curves
derived from Eq.3 are superimposed on the data in Fig.~1.

As can be seen, the observational relations between core and virial radii are
in good agreement with the respective theoretical relations, described via
eq.(3) and obtained with our ($\alpha=0.8$) profiles. This strongly supports
the choice of these profiles in the description of cluster density
distributions.

Moreover, it is possible to compare the measured $R_{v}$ with the corresponding
values $R_{v,comp}$ computed from the observed $R_{c}$ and the respective
apertures using Eq.3. $R_{v}$ correlates with $R_{v,comp}$ at a significance
level higher than 99.99~\%. For the virial radii the regression bisector line
(see e.g. Isobe et al., 1990) is
\begin{equation}
R_{v,0.75,comp}=0.02(\pm 0.04)+0.96(\pm 0.06)\cdot R_{v,0.75}.
\end{equation}
Fig.$\ $2 shows $R_{v,0.75,comp}$ versus $R_{v,0.75}$.
Obviously, it is also possible to compare the measured $R_{c}$ with the
corresponding values $R_{c,comp}$ computed from the observed $R_{v}$ and the
respective apertures using Eq.3, and the result is quite similar.

As a consequence of this analysis, since the estimate of $R_{v}$ is usually
more straightforward than that of $R_{c}$, one can evaluate $R_{c}$ when
$R_{v}$ and the corresponding aperture are known.

We also investigated the existence of a correlation between $R_v$ (and $R_c$)
and other optical cluster quantities, such as the Abell richness, the
morphological types by Bautz-Morgan and by Rood-Sastry, as reported in Table 1,
and the robust velocity dispersions computed as in GBGMM:
we found no strong evidence of correlation.

The comparison between optical and X-ray core radii has been extensively
discussed with controversial results (see, e.g., Lea, Silk \& Kellogg 1973;
Smith, Mushotzky \& Selermitos 1979, Abramopoulous \& Ku 1983). Detailed
optical analyses of few clusters (see, e.g., Ku et al. 1983, Henry \& Henriksen
1986) suggest that optical and X-ray core radii are comparable within
observational errors (see also David et al. 1990). We considered the X-ray core
radii, $R_{x}$, of Jones \& Forman (1984); the median value of $\alpha$ in
their fits is $\sim 0.7\pm 0.1$, which agrees with our optical $\alpha=0.8$.
The distribution function of $R_{x}$ does not differ (according to the KS-test)
from that of our $R_{c,0.75,\alpha=0.8}$, when the common subsample
of 18 clusters is considered. Fig.~3 shows these distributions.

\subsection{Radii Distributions}
Our cluster set is not volume-complete.
Therefore, in order to estimate the
cosmic distributions for our cluster core and virial radii, we should
normalize our observational distributions to a complete cluster sample. For
this purpose, we considered the Cluster Catalogue of Abell, Corwin \& Olowin
(1989; hereafter ACO), which is nominally complete for clusters with redshift
$z \leq 0.2$ and richness class $R \geq 1$.

In GBGMM and Biviano et al. (1993) we used the Abell richness
distribution to normalize the distributions of cluster velocity dispersions
and, respectively, of cluster virial masses. The normalization makes sense
because the richnesses is available for all ACO clusters, and both
velocity dispersion and virial mass correlates with richness.
Since there is no significant correlation
between cluster size and richness, it is no obvious that a similar
normalization would work in this case. On the other hand, the same lack
of correlation suggests that the cluster size distribution should not
depend critically on completeness. We decided to perform the normalization
to richness, anyway.

The normalization was accomplished
by randomly extracting 1000 values of $R_{c}$ and $R_{v}$ from the
corresponding values measured in our clusters, in such a way as to reproduce
the richness distribution of ACO's cluster sample (poor sample statistics
forced us to consider clusters with $R \geq 3$ as belonging to the same class).

We considered the values of $R_{c,0.75,\alpha=0.8}$, listed in Table 4,
having excluded the $R=0$ clusters (ACO is complete for $R \geq 1$). The
corresponding ACO-normalized cumulative distributions is shown in Fig.4;
the median value of this distribution is $0.17^{+0.04}_{-0.05}$\h (where the
99\% interval limits of the median are indicated).

Since the virial radii depend on the apertures used, we cannot provide a single
cosmic distribution of virial radii. So, we chose the aperture 0.75\h (half an
Abell radius) and the values of $R_{v,0.75}$  given in
Table~2. The ACO-normalized
cumulative distributions for $R_{v,0.75}$ are shown in Fig.5. The median
value (with 99\% interval limits) is $0.67^{+0.04}_{-0.06}$\h.

\section{Conclusions}
The cluster sample considered here is the most extensive one
 in the literature for
the evaluation of virial radii and the fitting of galaxy density
profiles, but it is not volume-complete.

Most of our cluster profiles ($\sim 90\%$) are fitted by
hydrostatic-isothermal-like profiles with $\alpha=0.8$; this result is,
therefore, quite general.

Our optical results are consistent with the hypothesis that many clusters,
within the errors and the apertures considered, are satisfactory  described by
isothermal models. In fact, the slope of our best-fit profiles is similar to
that of X-ray cluster profiles (e.g., Jones \& Forman 1984). Therefore we agree
with the analysis of Bahcall and Lubin (1993), who suggested a value of
$\alpha=0.7 \pm 0.1$ in order to resolve the $\beta$-discrepancy.

This result may seem surprising in view of the mounting evidence
that subclustering is ubiquitous in clusters (see, e.g., Bird 1994).
However, as discussed by West (1990), evidence for subclustering
does not necessarily imply dynamical youth for the cluster. Classical
tests for subclustering do not make any distinction between the case
of simple geometrical projection of a group onto a cluster field,
the case of a group that is falling into an otherwise relaxed cluster,
and the case of real substructure, i.e. a true perturbation of the
cluster dynamical potential. Moreover, subcluster masses are often
an order of magnitude lower than their parent cluster masses
(Escalera et al. 1994), and therefore their influence on the main
cluster properties need not to be huge.

Our virial radii correlate with our core radii at a very high significance
level. The relationship between these two size-estimates obeys the theoretical
relationship which links the virial radius to the density profile with
$\alpha=0.8$.

We provide the distribution functions of $R_{v}$ and $R_{c}$ obtained within a
fixed aperture of half an Abell radius. The analysis of the radii errors allows
us to claim that the dispersion of the values of the cluster radii is partially
intrinsic. The median value of our core radii is $0.17$\h and the median
value of our virial radii (within half an Abell radius) is $0.67$\h.

\vspace{0.5cm}
We thank Craig Sarazin for kindly providing us with his FORTRAN code for
the Maximum Likelihood analysis of density profiles, John Hill for
giving us an electronic copy of his cluster data,
and Michael West, the Referee, for his useful suggestions.
This work was partially supported by the
{\em Ministero per l'Universit\`a e per la Ricerca scientifica e tecnologica},
and by the {\em Consiglio Nazionale delle Ricerche (CNR-GNA)}.

\pagebreak
\section*{Appendix}
Both our core and  virial radii ($R_c$ and $R_v$, respectively) describe the
galaxy distribution inside the cluster, so the existence of a relation between
these two quantities is quite natural.
This functional dependence can be derived
when one knows the density profile.

The $3D$ virial radius $R_{v,3D}$ of a distribution of $N$ cluster galaxies of
masses $m_{i}$ on the sky plane,  within a circular aperture of radius $A$,  is
given  by the usual relation expressing the total potential energy of the
system.  Since only projected positions are known for the galaxies, a
de-projection factor (usually $\pi$/2) is used to deduce $R_{v,3D}$ from its
projected  value $R_{v}$, which is derived from the relation
$$-\frac{G}{R_v}\;\left({\sum_{1}^{N}}_{i} m_i\right)
^2=-G{\sum_{1}^{N}}_{i\neq j}\frac{m_im_j}{R_{ij}},\;\;\;\;(A1)$$
which becomes our Eq.(2) on the hypothesis that galaxy masses do not correlate
with galaxy positions in clusters. This hypothesis is largely assumed in the
literature and it is justified by the absence of pronounced luminosity
segregation in galaxy clusters (see, e.g., Biviano et al. 1992).

The relation (A1) describes the total potential energy of the $2D$ distribution
of masses. It is thus possible to evaluate $R_{v}$ for a given surface-density
profile $\sigma (r/R_{c})$ ($r \leq A$), knowing the total potential energy
$W_{2D}(A)$ of the disk of radius $A$. This energy is:
$$W_{2D}(A)=\frac 12\;2\pi \int_0^A\Phi (r)\;\sigma
(r/R_{c})\;r\;dr;\;\;\;\;(A2)$$
in (A2) the factor $1/2$ avoids counting twice the same couples of mass
elements in the disk, and the gravitational  potential $\Phi (r)$ (using polar
coordinates $(r^{\prime},\phi)$) is:
$$\Phi (r)=-G\int_0^A\int_0^{2\pi }\frac{\sigma
(r^{\prime }/R_c)\;r^{\prime }}{r^2+r^{\prime 2}-2r\;r^{\prime
}\;\cos \phi }\;d\phi \;dr^{\prime }=-G\;R_c\;I(r/R_c).\;\;\;\;(A3)$$
The function $I(r/R_{c})=I(x)$
(where $x = r/R_{c}$) may be expressed, using the
complete elliptical integral of first kind $K(y)$ (see Binney
and Tremaine, pg.73, 1987), as
$$I(x)=4\int_0^{A/R_c}\frac{\sigma (u)\;u}{x+u}\;K\left[
\frac{4\;u\;x}{\left( u+x\right) ^2}\right] \;du,\;\;\;\;(A4)$$
where $u=r^{\prime}/R_c$, and $K(y)$ is defined by
$$K(y)=\int_0^1\frac 1{\sqrt{\left( 1-t^2\right) \left(
1-y\;t^2\right) }}dt.\;\;\;\;(A5)$$

{}From these relations it is possible to obtain the relation
between $R_v$ and $R_c$, for a given profile and aperture:
$$R_v=R_c\frac{4\pi \left( \int_0^{A/R_c}\sigma
(x)\;x\;dx\right) ^2}{\int_0^{A/R_c}I(x)\;\sigma
(x)\;x\;dx}=R_c\;F (A/R_c).\;\;\;\;(A6).$$

The function $F(A/R_c)$
can be evaluated for any given surface density profile $\sigma (r/R_c)$.
For the profile
$$\sigma (r/R_c)=\frac{\sigma _0}{\left[ 1+\left( r/R_c\right)
^2\right] ^{0.8}}.\;\;\;\;(A7)$$
$F(A/R_c)$ is well represented, for
$0.01 \leq (A/R_{c}) \leq 100$ and with an accuracy better than 3\%,
by the relation
$$F(A/R_c)= 1.193\;(A/R_{c})\;\frac{1+ 0.032\;(A/R_{c})}{1+ 0.107\;(A/R_{c})}.
\;\;\;\;(A8)$$

\pagebreak
\section*{References}

\ref Abell, G.O., Corwin, H.G.Jr., \& Olowin, R.P. 1989, ApJS, 70, 1 (ACO)

\ref Abramopoulous, F., \& Ku, W.H. 1983, ApJ, 271, 446

\ref Avni, Y. 1976, ApJ, 210, 642

\ref Bahcall, N.A. 1975, ApJ, 198, 249

\ref Bahcall, N.A. 1977, ARA\&A, Vol.15, 505

\ref Bahcall, N.A., \& Lubin, L.M. 1993, ApJ in press

\ref Beers, T.C., Flynn, K., \& Gebhardt, K. 1990, AJ, 100, 32

\ref Beers, T.C., Forman, W., Huchra, J.P., Jones, C., \&
Gebhardt, K. 1991, AJ, 102, 1581

\ref Beers, T.C., Gebhardt, K., Huchra, J.P., Forman, W., Jones, C.,
\& Bothun, G.D. 1992, ApJ, 400, 410

\ref Beers, T.C., Geller, M.J., Huchra, J.P., Latham, D.W., \& Davis, R.J.
1984, ApJ, 283, 33

\ref Bell, M., Whitmore, B.C. 1989, ApJS, 70, 139

\ref Binggeli, B., Sandage, A., \& Tammann, G.A. 1985, AJ, 90, 1681

\ref Binney, J., \& Tremaine, S. 1987, {\em Galactic Dynamics},
Princeton Univ. Press

\ref Biviano, A., Girardi, M., Giuricin, G., Mardirossian, F., \&
Mezzetti, M. 1992, ApJ, 396, 35

\ref Biviano, A., Girardi, M., Giuricin, G., Mardirossian, F., \&
Mezzetti, M. 1993, ApJ, 411, L13

\ref Bothun, G.D., Aaronson, M., Schommer, B., Mould, J., Huchra, J.,
\& Sullivan, W.T. III 1985, ApJS, 57, 423

\ref Bothun, G.D., Geller, M.J., Beers, T.C.,  \& Huchra, J.P. 1983,
ApJ, 268, 47

\ref Bowers, R.G., Ellis, R.S., \& Efstathiou, G. 1988, MNRAS, 234, 725

\ref Cavaliere, A., Colafrancesco, S., \& Scaramella, R. 1991, ApJ, 380, 15

\ref Chapman, G.N.F., Geller, M.J.,  \& Huchra, J.P. 1987, AJ, 94, 571

\ref Chapman, G.N.F., Geller, M.J.,  \& Huchra, J.P. 1988, AJ, 95, 999

\ref Chincarini, G.,  \& Rood, H.J. 1976, PASP, 88, 388

\ref Colless, M.,  \& Hewett, P. 1987, MNRAS, 224, 453

\ref Cristiani, S., de Souza, R., D'Odorico, S., Lund, G.,  \&
   Quintana, H. 1987, A\&A, 179, 108

\ref David, L.P., Arnaud, K.A., Forman, W., \& Jones, C. 1990, ApJ, 356, 32

\ref Dickens, R.J., Currie, M.J., \& Lucey, J.R. 1986, MNRAS, 220, 679

\ref Dickens, R.J.,  \& Moss, C. 1976, MNRAS, 174, 47

\ref Dressler, A. 1978, ApJ, 226, 55

\ref Dressler, A., \&  Shectman, S.A. 1988b, AJ, 95, 284

\ref Dressler, A., \&  Shectman, S.A. 1988a, AJ, 95, 985

\ref Durret, F., Gerbal, D., Lachi\`eze-Rey, M., Lima-Neto, G., Sadat, R.
1993, A\&A in press

\ref Edge, A.C., \& Stewart, G.C. 1991, MNRAS, 252, 428

\ref Efron, B., 1979, S.I.A.M. Rev, 21, 460

\ref Escalera, E., Biviano, A., Girardi, M., Giuricin, G., Mardirossian, F.,
Mazure, A., \& Mezzetti, M. 1994, ApJ, 423, 539

\ref Eyles, C.J., Watt, M.P., Bertram, D., Church, M.J., Ponman, T.J.,
  Skinner, G.K., \& Willmore, A.P. 1991, ApJ, 376, 23

\ref Faber, S.M.,  \& Dressler, A. 1977, AJ, 82, 187

\ref Fabricant, D.G., Kent, S.M.,  \& Kurtz, M.J. 1989, ApJ, 336, 77

\ref Fabricant, D., Kurtz, M., Geller, M., Zabludoff, A., Mack, P.,  \&
   Wegner, G. 1993, AJ, 105, 788

\ref Fernley, J.A., \&  Bhavsar, S.P. 1984, MNRAS, 210, 883

\ref Fitchett, \& Merritt, D. 1988, ApJ, 335, 18

\ref Gavazzi, G. 1987, ApJ, 320, 96

\ref Geller, M.J. 1984, in {\em Clusters and Groups of Galaxies},
F. Mardirossian, G. Giuricin, \& M.Mezzetti Eds., D. Reidel Pub. Com.
,Dordrecht, Holland, p. 353

\ref Geller, M.J., Beers, T.C., Bothun, G.D., \& Huchra, J.P. 1984, AJ, 89, 319

\ref Gerbal, D., Durret, F., Lachi\`eze-Rey, M. 1994, A\&A, in press

\ref Gerbal, D., Durret, F., Lima-Neto, G., \& Lachi\`eze-Rey, M. 1992, A\&A,
 253, 77

\ref Giovanelli, R., Haynes, M.P.,  \& Chincarini, G.L. 1982, ApJ, 262, 442

\ref Girardi, M., Biviano, A., Giuricin, G., Mardirossian, F., \&
Mezzetti, M. 1993, ApJ, 404, 38 (GBGMM)

\ref Giuricin, G., Mardirossian, F., \& Mezzetti, M. 1984, A\&A,
141, 419.

\ref Gregory, S.A.,  \& Thompson, L.A. 1978, ApJ, 222, 784

\ref Gregory, S.A.,  \& Thompson, L.A. 1984, ApJ, 286, 422

\ref Gregory, S.A., Thompson, L.A.,  \& Tifft, W.G. 1981, ApJ, 243, 411

\ref Henry, J.P., \& Henriksen, M.J. 1986, ApJ, 301, 689

\ref  Hill, J.M., \& Oegerle, W.R. 1993, AJ, 106, 831

\ref Hintzen, P., Hill, J.M., Lindley, D., Scott, J.S.,  \&
 Angel, J.R.P. 1982, AJ, 87, 1656

\ref Hintzen, P., Oegerle, W.R.,  \& Scott, J.S. 1978, AJ, 83, 478

\ref Inagaki, S., Itoh, M., \& Saslaw W.C. 1992, ApJ, 386, 9

\ref Isobe, T., Feigelson, E.D., Akritos, M.G., \& Babu, G.J. 1990,
   ApJ, 364, 104

\ref Jones, C., \& Forman, W. 1984, ApJ, 276, 38

\ref Jones, C., \& Forman, W. 1992, in {\em Clusters and Superclusters of
Galaxies}, A.C. Fabian ed., Kluwer Academic Pub., The Netherlands, p.49

\ref Kashlinsky, A. 1992, ApJ, 386, L37

\ref Kent, S.M.,  \& Gunn, J.E. 1982, AJ, 87, 945

\ref Kent, S.M.,  \& Sargent, W.L.W. 1983, AJ, 88, 697

\ref King, I.R. 1962, AJ, 67, 471

\ref King, I.R. 1966, AJ, 71, 64

\ref Kneib, J.-P., Mellier, Y., Fort, B., Mathez, G. 1993, A\&A, 273, 367

\ref Ku, W.F., Abramopoulos, F., Nulsen, P.E.J., Fabian, A.C.,
Stewart, G.C., Chincarini, G., Tarenghi, M. 1983, MNRAS, 203, 253

\ref Kurtz, M.J., Huchra, J.P., Beers, T.C., Geller, M.J., Gioia, I.M.,
      Maccacaro, T., Schild, R.E.,  \& Stauffer, J.R. 1985, AJ, 90, 1665

\ref Lauberts, A.,  \& Valentjin, E.A. 1989,
{\em The Surface Photometry Catalogue of the ESO-Uppsala Galaxies},
Garching bei Muenchen: ESO

\ref Lea, S., Silk, J., Kellogg, E., \& Murray, S. 1973, ApJ, 184, L105

\ref Lilje, P.B., \& Lahav, O. 1991, ApJ, 374, 29

\ref Limber, D.N.,  \& Matthews, W.G. 1960, ApJ, 132, 286

\ref  Lubin, L.M., \& Bahcall, N.A. 1993, ApJ, 415, L17

\ref Lynden-Bell, D. 1967, MNRAS, 136, 101

\ref Malumuth, E.M., \& Kriss, G.A. 1986, ApJ, 308, 10

\ref Malumuth, E.M., Kriss, G.A., Van Dyke Dixon, W., Ferguson, H.C.,
\& Ritchie, C. 1992, AJ, 104, 495

\ref Metcalfe, N., Godwin, J.G.,  \& Spenser, S.D. 1987, MNRAS, 225, 581

\ref Miralda-Escud\'e, J. 1992, ApJ, 390, L65

\ref Moss, C.,  \& Dickens, R.J. 1977, MNRAS, 178, 701

\ref Oegerle, W.R., \& Hill, J.M. 1993, AJ, 104, 2078

\ref Oemler, A.Jr. 1974, ApJ, 194, 1

\ref Ostriker, E.C., Huchra, J.P., Geller, M.J., \& Kurtz, M.J. 1988, AJ,
96, 177

\ref Peebles, P.J. 1970, AJ, 75, 13

\ref Peebles, P.J. 1976, ApJ, 205, L109

\ref Peebles, P.J., Daly, R.A.,  \&  Juszkiewicz, R. 1989, ApJ, 347, 563

\ref Perea, J., Del Olmo, A., \& Moles, M. 1990, A\&A, 237, 319

\ref  Pisani, A., Giuricin, G., Mardirossian, F., \& Mezzetti, M.
1992, ApJ, 389, 68

\ref Postman, Huchra, J.P., M.,  \& Geller, M.J. 1986, AJ, 92, 1238

\ref Proust, D., Quintana, H., Mazure, A., da Souza, R., Escalera, E.,
Sodr\`e, L.Jr.,  \& Capelato, H.V. 1992, A\&A, 258, 243

\ref Quintana, H., Melnick, J., Infante, L.,  \& Thomas, B. 1985, AJ, 90, 410

\ref Quintana, H.,  \& Ramirez, A. 1990, AJ, 100, 1424

\ref Rhee, G., van Harleem, M., \& Katgert, P. 1992, AJ, 103, 1721

\ref Richter, O.G., 1987, A\&AS, 67, 237

\ref Richter, O.G., 1989, A\&AS, 77, 237

\ref Richter, O.G.,  \& Huchtmeier, W.K. 1982, A\&A, 109, 155

\ref Sarazin, C.L. 1980, ApJ, 236, 75

\ref Sarazin, C.L. 1986, Rev.Mod.Phys., 58, 1

\ref Sharples, R.M., Ellis, R.S.,  \& Gray, P.M. 1988, MNRAS, 231, 479

\ref Smith, H. Jr. 1980, ApJ, 241, 63

\ref Smith, B.W., Mushotzky, R.F., \& Serlemitos, P.J. 1979, ApJ, 227, 37

\ref Sodr\`e, L., Capelato, H.V., Steiner, J.E., Proust, D.,  \& Mazure, A.
     1992, MNRAS, 259, 233

\ref Soucail, G. 1992, in {\em Clusters and Superclusters of Galaxies},
A.C. Fabian ed., Kluwer Academic Pub., The Netherlands, p.199

\ref Stauffer, J., Spinrad, H.,  \& Sargent, W.L.W. 1979, ApJ, 228, 379

\ref Stepanyan, Dzh.A. 1984,  Astrophysics, 20, 478

\ref Tarenghi, M., Tifft, W.G., Chincarini, G., Rood, H.J.,  \& Thompson, L.A.,
      1979, ApJ, 234, 793

\ref Teague, P.F., Carter, D.,  \& Gray, P.M. 1990, ApJS, 72, 715

\ref Tifft, W.G. 1978, ApJ, 222, 54

\ref Tully, R.B. 1988, {\em Nearby Galaxy Catalog,} Cambridge Univ. Press

\ref West, M.J. 1990, in {\em Clusters of Galaxies,} W.R.Oegerle \&
M.J.Fitchett Eds., Cambridge Univ.Press

\ref West, M.J.,  Dekel, A., \& Oemler, A., 1987, ApJ, 316, 1

\ref West, M.J., Oemler, A., \& Dekel, A. 1989, ApJ, 327, 1

\ref Willmer C.N.A., Focardi P., Chan R., Pellegrini P.S.,  \& Nicolaci Da
      Costa L. 1991, AJ, 101, 57

\ref Willmer C.N.A., Focardi P., Nicolaci Da Costa L.,  \& Pellegrini P.S.
      1989, AJ, 98, 1531

\ref Zabludoff, A.I., Huchra, J.P.,  \& Geller, M.J. 1990, ApJS, 74, 1

\pagebreak

\section*{Captions to the Figures}
\subsection*{Fig.$\ $1:}
$\log R_{c,0.5,\alpha=0.8}$ vs. $\log R_{v,0.5}$ (panel a), and
$\log R_{c,0.75,\alpha=0.8}$ vs. $\log R_{v,0.75}$ (panel b).
The superimposed curves are derived from Eq.3.
\subsection*{Fig.$\ $2:}
$R_{v,comp}$ versus $R_{v}$. The superimposed regression line (Eq.4) does not
significantly differ from the bisector line.
\subsection*{Fig.$\ $3:}
The cumulative distributions of $R_{c,0.75,\alpha=0.8}$ (solid line) and
of the X-ray core radii of Jones \& Forman (1984) for the common
subsample of clusters.
\subsection*{Fig.$\ $4:}
The ACO-normalized cumulative distribution of the core radii.
\subsection*{Fig.$\ $5:}
The ACO-normalized cumulative distribution of the virial radii,
evaluated within half an Abell radius.

\pagebreak
\section*{Table 1 (continued)}
\begin{description}
\item{Col.(1): } Cluster name;
\item{Col.(2): } Number of member galaxies
within 1.5 $h^{-1}$ Mpc from the cluster centre;
\item{Col.(3): } Abell number counts, C;
\item{Col.(4): } Richness class, R;
\item{Col.(5): } Bautz-Morgan type;
\item{Col.(6): } Roods-Sastry type;
\item{Col.(7): } Probability of absence of substructure;
\item{Col.(8): } Reference to velocities.
\end{description}
\small{
[1] Malumuth et al. (1992);
[2] Fabricant et al. (1993);
[3] Proust et al. (1992);
[4] Hill $\&$ Oegerle (1993);
[5] Chapman, Geller, $\&$ Huchra (1988);
[6] Giovanelli et al. (1982);
[7] Gregory, Thompson, $\&$ Tifft (1981);
[8] Moss $\&$ Dickens (1977);
[9] Beers et al. (1992);
[10] Kent $\&$ Sargent (1983);
[11] Colless $\&$ Hewett (1987);
[12] Quintana $\&$ Ramirez (1990);
[13] Ostriker et al. (1988);
[14] Dressler $\&$ Shectman (1988b)
[15] Hintzen et al. (1982);
[16] Stepanyan (1984);
[17] Kurtz et al. (1985);
[18] Zabludoff, Huchra, $\&$ Geller (1990);
[19] Hintzen, Oegerle, $\&$ Scott (1978);
[20] Beers et al. (1991);
[21] Chapman, Geller, $\&$ Huchra (1987);
[22] Richter (1987);
[23] Richter (1989);
[24] Geller et al. (1984);
[25] Teague, Carter, $\&$ Gray (1990);
[26] Gavazzi (1987);
[27] Gregory $\&$ Thompson (1978);
[28] Tifft (1978);
[29] Dickens $\&$ Moss (1976);
[30] Kent $\&$    Gunn (1982);
[31] Faber  $\&$ Dressler (1977);
[32] Bowers, Ellis,  $\&$  Efstathiou (1988);
[33] Quintana et al. (1985);
[34] Postman, Huchra,  $\&$  Geller (1986);
[35] Oegerle  $\&$ Hill (1993);
[36] Tarenghi et al. (1979);
[37] Gregory  $\&$ Thompson (1984);
[38] Stauffer, Spinrad, $\&$  Sargent (1979);
[39] Fabricant, Kent,  $\&$ Kurtz (1989);
[40] Sharples, Ellis, $\&$ Gray (1988);
[41] Dickens, Currie, $\&$ Lucey (1986);
[42] Lauberts $\&$ Valentjin, E.A. (1989);
[43] Cristiani et al. (1987);
[44] Metcalfe, Godwin,  $\&$ Spenser (1987);
[45] Willmer et al. (1991);
[46] Sodre' et al. (1992);
[47] Beers et al. (1984);
[48] Bothun et al. (1983);
[49] Willmer et al. (1989);
[50] Malumuth  $\&$ Kriss (1986);
[51] Richter  $\&$ Huchtmeier (1982);
[52] Bothun et al. (1985);
[53] Chincarini  $\&$ Rood (1976);
[54] Bell  $\&$ Whitmore (1989);
[55] Tully (1988);
[56] Binggeli, Sandage  $\&$ Tammann (1985).

}
\end{document}